# Citation as a Representation Process

Victor V. Kryssanov[a1,], Frank J. Rinaldo[a], Hitoshi Ogawa[a] and Evgeny L. Kuleshov[b]
[a] *Faculty of Engineering and Information Science, Ritsumeikan University, Kusatsu, JAPAN*
[b] *Department of Computer Systems, the Far-Eastern National University, Vladivostok, RUSSIA*

The presented work proposes a novel approach to model the citation rate. The paper begins with a brief introduction into informetrics studies and highlights drawbacks of the contemporary approaches to modeling the citation process as a product of social interactions. An alternative modeling framework based on results obtained in cognitive psychology is then introduced and applied in an experiment to investigate properties of the citation process, as they are revealed by a large collection of citation statistics. Major research findings are discussed, and a summary is given.

## 1 INTRODUCTION

A recently increasingly popular area of scientific enquiry called informetrics deals with citation analysis by investigating statistical properties of the process of referring to related work by authors of scientific and technical articles. Informetrics endeavors to establish the social mechanisms underlying the creation and propagation of knowledge in a research community that would allow for formulation of universal principles for objective evaluation of individual contributions in the knowledge development process. The latter is of utmost importance to institutions involved in various resource distribution activities.

In his seminal work on citation statistics published in 1965 [8], D.J.S. Price suggested the cumulative advantage stochastic mechanism as the driving force of the citation process that, in fact, determined the research areas and directions of informetrics for the following four decades. The key conjecture of the classic model is that the observed (ir)regularities in the production, distribution, and consumption of information commodities (e.g. in the citation process) can be explained in terms of simple behavioral patterns – *social interactions* (e.g. preferential choice) prevailing among the members of a social group – that result in an apparently scale-free collective behavior (i.e. the power-law). Despite the huge popularity of the original idea, which has become the cornerstone of numerous social network theories exploiting the power law hypothesis, there is yet no support for it other than that obtained via simulation experiments, where the principal argument is typically built around whether a histogram of the artificial data resembles, as to an eye, the one of a specific data set at hand. Given the poor predictive capability of the classical model (see Fig.1 A and B), this inevitably raises questions both about its statistical soundness and the meaning of the model parameters. Furthermore, as many recent reports (e.g. Ref. [9]) suggest that in reality, up to 90% of citations in a paper are simply copied from other publications, about 10% are self-citations, and 7% are erroneous, the very idea of considering citations as a product of social interactions becomes problematic.

This note presents an attempt to develop an alternative approach to study informetrics phenomena. The underlying motivation is to arrive at such a theoretical model that *a*) all or at least some of its characteristics can be validated with procedures different from model simulation, and *b*) a thorough statistical analysis, different from least-square linear fitting of histograms on the double-logarithmic scale, does not defy the model forecast.

## 2 GENERAL FRAMEWORK

The main premise of our theory is that citation, considered as an information process, does not differ in principle from other modes of communication. The material of this section is therefore in line with the communication model originally proposed in Ref. [4].

The citation process can be viewed as a mechanism for substantiating the key idea(s) (concept, problem, theory, statement, etc.) of a paper. In general, it is the foundation upon which arguments are based, assertions are asserted, and basic ideas are left unexplained. It is up to the reader to fully explore the foundations (citations) of the new idea or concept present in a given text. We will thus assume that every instance of citing an article indicates the act of signification (by the citation) of a concept or idea otherwise not presented (e.g. owing to space limitations) in the main text with other media, such as words, formulas, diagrams, and the like. We will also assume that the citing results from cognitive processing (problem solving or decision-making) associated with the signified idea. The processing time $\tau$ – the time for which a problem remains unsolved and/or receives attention – depends on many factors, such as complexity of the underlying concept, current "fashion" in the field, results obtained in related domains, etc. and generally exceeds the time-frame allocated for writing a particular text. Both the cited articles and concepts/ideas behind them are shared by individual "citers;" we will consider a statistical ensemble of such individuals.

---

[1] Corresponding Author: Victor V. Kryssanov. Faculty of Engineering and Information Science, Ritsumeikan University, 1-1-1 Nojihigashi, Kusatsu, Shiga, 525-8577 JAPAN. E-mail: kvvictor@is.ritsumei.ac.jp

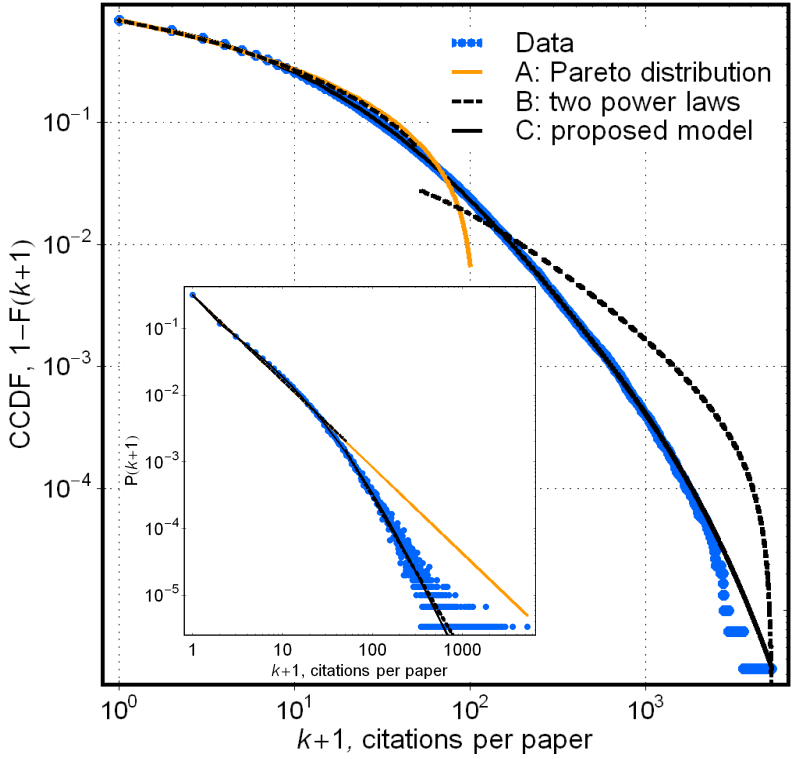

Fig.1. Statistical modeling of the citation process: A large data sample comprising citation information (for details on the data, see Section 3.2) was fitted to three different models. It is obvious from the complementary cumulative distribution function (CCDF) plots that neither a single Pareto distribution (A) nor a linear combination of two power laws (B) recently suggested by the authors of Ref. [6] provides for a reasonably accurate fit. In contrast, the predictive power of the 5-parameter bimodal form proposed in this study is remarkable. Arguments like "it is too complex to be true," while expected, are similar to claims that "since a majority of people speak Chinese, it must be the simplest language."
Inset: The same results presented with the corresponding probability density functions (histograms) are, however, misleading, as cases B and C appear nearly indistinguishable.

Let us denote $K_0$ a discrete random variable indicating the count of different citations representing one fixed concept (no matter how complex). It appears natural to assume that values of $K_0$ will be determined by the concept processing time $\tau$: for any fixed observation time, the longer, on average, the processing time, the more frequent the citation (an article with zero citations corresponds to zero processing time, e.g. when the considered concept/problem is outdated, notorious, "uninteresting," remains unnoticed, or is still poorly understood by anybody other than the article authors). The latter means that there is a positive correlation between a number of citations that a particular problem receives and the time for which this problem remains in the community's focus. The same citation may stand (be used) for different concepts, and the same concept may be signified with different citations.

To characterize the behavior of $K_0$, we will first seek to estimate $f_{K_0}(s)$, $s = 1, 2, \ldots$, its probability mass function, PMF (for technical reasons, we define the domain of the distribution as strictly positive, i.e. $s > 0$). Due to Edwin T. Jaynes, it is well-known that the "the least biased estimate possible on the given information" [3] is the distribution that maximizes the Shannon entropy $H = -\sum_s f_{K_0}(s) \ln f_{K_0}(s)$ subject to the normalization $\sum_s f_{K_0}(s) = 1$ and expectation $\sum_s s f_{K_0}(s) = \bar{k}_0$ constraints; here $\bar{k}_0$ is the expectation of $K_0$. The Lagrangian for this optimization problem is given by the following expression:

$$\mathcal{L}(f_{K_0}) = -\sum_s f_{K_0}(s) \ln f_{K_0}(s) + \gamma(1 - \sum_s f_{K_0}(s)) + \beta(\bar{k}_0 - \sum_s s f_{K_0}(s)), \qquad (1)$$

where the $\gamma$ and $\beta$ terms are Lagrangian multipliers. The optimality conditions $\partial \mathcal{L} / \partial f_{K_0}(s) = 0$ and $\partial \mathcal{L} / \partial \gamma = 0$ allow us, after some algebra, to obtain the probability function sought:

$$f_{K_0}(s) = \Pr[K_0 = s \mid \beta] = (e^\beta - 1) e^{-s\beta}, \quad s = 1, 2, \ldots. \qquad (2)$$

$\bar{k}_0 = \sum_{s=1}^\infty s f_{K_0}(s) = e^\beta / (e^\beta - 1) = 1 / f_{K_0}(1)$, then if $\bar{k}_0 \gg 1$, it follows that $f_{K_0}(1) \ll 1$ and $\beta \ll 1$, hence

$\beta \approx 1/\bar{k}_0$ and is a stochastic parameter.

Let us next consider a more realistic situation when a measured stochastic variable $K$ indicates citation occurrences for not just one but many and different concepts. The statistical properties of $K$ will then depend on the parameter $\beta$ that can naturally vary (e.g. as a result of a variation in the cognitive processing time). For the sake of simplicity, we will consider $\beta$ as a continuous random variable. Let $g(\beta)$ be the probability density function (PDF) of $\beta$. If the number of concepts/problems signified with citations is large, $f_K(s)$, $s = 1, 2, \ldots$, the PMF of $K$ can be obtained as a $g(\beta)$ parameter-mix of $f_{K_0}(s | B = \beta)$:

$$f_K(s) = f_{K_0}(s | B = \beta) \bigwedge_B g(\beta) = \int_0^\infty f_{K_0}(s) g(\beta) d\beta . \qquad (3)$$

Finally, it is important to understand that parameter $\beta$ varies as the inverse of $\bar{k}$ the average citation occurrence that, in turn, is affected by the processing time $\tau$ associated with the investigated (via citations) class of problems (ideas, theories, etc). As the problems may not necessarily be statistically homogeneous in terms of the processing time, there can exist several distributions for $g(\beta)$. A generalized form of Equation (3) for $P(k)$ the PMF of the random variable $K$ can then be expressed as follows:

$$P(k) = \sum_{i=1}^M c_i f_{i_K}(k), \quad k = 1, 2, \ldots, \qquad (4)$$

where weights $0 < c_i \le 1$ give the likelihood to observe the influence of the $i$-th type problems, as in terms of $\tau$ the time expended for their solution (consideration, discussion, or analysis), on the random variable $K$, and each $f_{i_K}(k)$ is defined with Equation (3).

## 3 STATISTICAL PROPERTIES OF THE CITATION PROCESS

### 3.1 The Model

The modeling framework of the previous section stipulates that to correctly describe the citation process, it is necessary to adequately characterize the dynamics of problem-solving that takes place in the community and is a cause of citation. In other words, we need to define an appropriate distribution function for the time $\tau$ a problem receives (i.e. is represented with) citations. Based on the earlier formulated condition that $g(\beta) \overset{d}{=} f(1/\tau)$ (here, "$\overset{d}{=}$" means "distributionally equivalent" and may, in certain situations, also be read as "behaves as"), the latter will allow for deriving a functional form for the distribution (4).

The most general facts about human decision-making, problem-solving, and similar activities dictate that the sought distribution must be defined on the non-negative domain extended to positive infinity (assuming that some problems are never solved) and the probability of $\tau = 0$ should be close to 0 (of course, if assuming that no problems are solved instantaneously). In terms of the hazard rate $h(t) = \frac{f(t)}{1-F(t)}$, which is usually used to classify distributions ($f(t)$ and $F(t)$ are the PDF and the cumulative probability function, CDF of a distribution, respectively), it is natural to expect that for the cognitive processes, $h(\tau)$ should first rise from (nearly) zero and then decrease to a non-zero value as $\tau \to \infty$. It has been shown [2] that among the whole variety of the relevant distributions utilized in cognitive sciences, the best candidate is the Wald (also called Inverse Gaussian) distribution with a PDF $f(\tau) = \sqrt{\frac{\lambda}{2\pi\tau^3}} e^{-\frac{\lambda(\tau-\mu)^2}{2\tau\mu^2}}$, where $\lambda, \mu > 0$ are the location and the scale parameters, respectively; for this PDF, the expectation $E(\tau) = \mu$, and the variance $Var(\tau) = \mu^3/\lambda$.

It is now quite straightforward to derive:

$$g(\beta) \overset{d}{=} f(1/\tau) = \sqrt{\frac{\lambda}{2\pi\tau}} e^{-\frac{\lambda(\tau\mu-1)^2}{2\tau\mu^2}}, \qquad (5)$$

where parameters $\lambda, \mu > 0$, while $E(\beta) = \mu^{-1} + \lambda^{-1}$ and $Var(\beta) = (2\mu + \lambda)/(\mu\lambda^2)$. (As it has been discussed in Section 2, $1/\beta$ may be interpreted as an estimate of the average citation *relative* rate.)

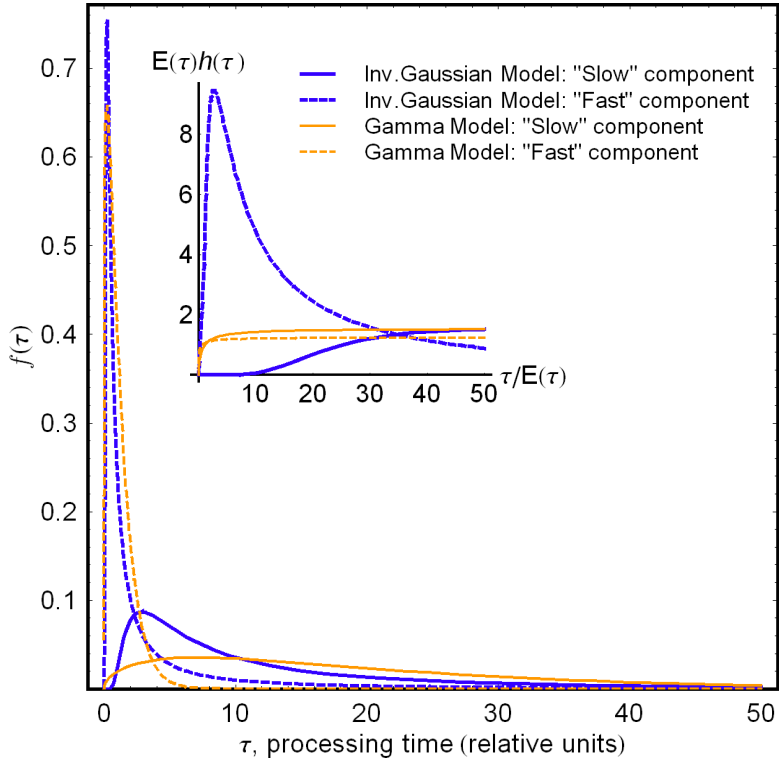

Fig.2. Characteristic patterns in the time devoted to a problem in the research community as revealed by citations in high-energy physics papers.
Inset: The hazard (decision-making or problem-solving) rate calculated for the distribution functions shown in the main figure. In the given context, the hazard rate $h(\tau)$ is essentially the rate at which a problem is solved or/and simply becomes uninteresting from the standpoint of citation at time $\tau$ given that the problem has not been solved at an earlier time.
Light lines (main figure and inset): the same data fitted to a model with a gamma-distributed processing time.
The displayed probability distribution forms are both extensively used in cognitive psychology for modelling the response/reaction time (Gamma) and decision making time delay (Wald/Inverse Gaussian) [7].

The direct substitution of Equation (5) into Equations (3-4) produces:

$$P_{WE}(k) = \sum_{i=1}^{M} c_i \, e^{\lambda_i / \mu_i} \sqrt{\lambda_i} \left( \frac{e^{-\sqrt{\frac{\lambda_i(2k-2+\lambda_i)}{\mu_i^2}}}}{\sqrt{2k-2+\lambda_i}} - \frac{e^{-\sqrt{\frac{\lambda_i(2k+\lambda_i)}{\mu_i^2}}}}{\sqrt{2k+\lambda_i}} \right), \quad k = 1,2,... \qquad (6)$$

that is hence the probability mass function of the citation occurrence number (*WE* stands for "Wald-Exponential").

### 3.2   Experiment

A dataset containing citation information about 299239 published journal papers in high energy physics accumulated in the SLAC SPIRES database since 1962 and obtained from the authors of Ref. [6] was used in an experiment conducted to explore the appropriateness of the assumptions formulated in Section 2 and validate the proposed model. The data collection appears reasonably homogeneous, as it represents the professional activities of a comparatively small research community, and it provides citation statistics for both cited and uncited articles (for details on the dataset, consult the original work [6]). The data were fitted to the model specified with Equation (6). Fig.1 (C) presents results of the modeling. $\hat{c}_i$, $\hat{\mu}_i$, and $\hat{\lambda}_i$ estimates for the corresponding parameters of the model were calculated using a numerical maximum likelihood method for $M$ running from 1 to 3. The equation with $M = 2$ was chosen for the modeling, because it yielded the smallest value of Akaike's Information Criterion, $\text{AIC} = -2\log(L(\hat{\phi}|k)) + 2n$, where $\log(L(\hat{\phi}|k))$ is the log-likelihood maximized with $\hat{\phi} = \{\hat{c}_{i \neq 1}, \hat{\mu}_i, \hat{\lambda}_i\}$, $i = 1,...,M$, and $n$ is the number of free parameters in the model). AIC is a measure assessing the relative Kullback-Leibler distance between the fitted model and the unknown true mechanism, which actually generated the observed data [1]. For the given dataset, Pearson's $\chi^2$ test does not reject the chosen model with a significance level $\alpha = 0.1$.

### 3.3   Discussion

It should be noted first, that while none of the citation mechanism models known to the authors from high-profile periodicals provide for a statistically sound fit to the entire range of the raw (i.e. without *any* preprocessing) data, the proposed 5-parameter distribution gives amazingly precise predictions for the citation rate. To explore whether the formula does not merely interpolate (i.e. overfits) the data, two probability distributions $f(\tau)$ of the cognitive processing time were reconstructed with

$\hat{\mu}_1 = 15.66$, $\hat{\mu}_2 = 11.72$, $\hat{\lambda}_1 = 8.92$ and $\hat{\lambda}_2 = 0.64$ the parameter values used (together with $\hat{c}_2 = 0.44$) for the fit shown in Fig.1 (C). As indicated in Fig.2 and its inset, the fitted model unveils two distinct patterns in the Wald-distributed processing time: one associated with frequently cited papers (solid lines), and another – with papers which received few or no citations at all (dashed lines). Papers with the "slow" pattern (i.e. complex and long-living ideas) got, on average, 10 times more citations (as estimated via $\beta \propto E(\tau)$) than papers with the "fast" processing time. This result – the two characteristic and quite natural regimes in human problem-solving – could hardly be detected using other models of the citation rate. For example, Fig.2 also displays the functions reconstructed from the same data using a two-component Lomax mixture model, which assumes a two-parameter Gamma distribution for $\tau$. This model is defined as

$$P_{GE}(k) = \sum_{i=1}^{2} c_i \left( \frac{b_i^{v_i}}{(k-1+b_i)^{v_i}} - \frac{b_i^{v_i}}{(k+b_i)^{v_i}} \right), k = 1, 2, \ldots, \quad (7)$$

where parameters $b_i, v_i > 0$, and it provides for a reasonably accurate fit (much better than any of the power-law models) to the citation data. Equation (7) has been proven a fairly universal model for communication processes [4]. This model is, however, unable to uncover the different citation regimes: the corresponding hazard rate curves are practically indistinguishable and uninformative (Fig.2, inset). The latter appears to be an important argument in support of our claim that Equation (6) does describe the citation mechanisms but is not just a "luckily guessed" interpolating formula (as is, for instance, Equation (7)).

The continuous analog of the PMF (6) is expressed as

$$f_{WE}(x) = \frac{\lambda\sqrt{2x+\lambda} + \mu\sqrt{\lambda}}{\mu(2x+\lambda)^{3/2}} e^{\frac{\lambda - \sqrt{\lambda(2x+\lambda)}}{\mu}}, \quad (8)$$

where $x$ is the measured stochastic variable, and is a version of the stretched exponential distribution – not really a "stranger" in informetrics studies (e.g. see [5]). It reveals two different modes, an exponential and a power-law with an exponent 1.5: $f_{WE}(x) \to \mu^{-1} e^{-x/\mu}$ as $\lambda \to \infty$, but $f_{WE}(x) \to \sqrt{\lambda}(2x+\lambda)^{-3/2}$ as $\mu \to \infty$. A detailed analysis of this asymptotic behavior is, however, left for future studies.

## 4 SUMMARY

The presented work offers one novel contribution – a possible explanation of the citation mechanisms from the positions of cognitive psychology using the apparatus of statistical physics. Another contribution would be the theoretically derived, not simply guessed, distribution form (6) that demonstrated to be an accurate, robust (this property was explored with diverse data samples not presented in this note) and, perhaps most importantly, meaningful model for the article citation rate.

### ACKNOWLEDGMENT

The authors thank Sune Lehmann for the data collection used in the presented study.

### REFERENCES

[1] Akaike, H. 1983. Information measures and model selection. International Statistical Institute, 44, pp.277-291.
[2] Fewell, M.P. 2004. Comparative Descriptive Statistics of Skewed Probability Distributions, Technical report DSTO-TR-1596. Australian Government, Department of Defense (DSTO), 37p.
[3] Jaynes, E.T. 1957. Information theory and statistical mechanics. Physical Review 106(4), pp.620-630.
[4] Kryssanov, V.V., Kakusho, K., Kuleshov, E.L., Minoh, M. 2005. Modeling hypermedia-based communication. Information Sciences 174(1-2), pp.37-53.
[5] Laherrere, J., Sornette, D. 1998. Stretched exponential distributions in nature and economy: "fat tails" with characteristic scales. The European Physical Journal B 2(4), pp.525-539.
[6] Lehmann, S., Lautrup, B., Jackson, A.D. 2003. Citation networks in high energy physics. Physical Review E 68(2), pp.026113-1-8.
[7] Luce, R.D. 1986. Response Times: Their Role in Inferring Elementary Mental Organization. New-York: Oxford University Press.
[8] Price, D.J.S. 1965. Networks of scientific papers. Science 149(3683), pp.510–515.
[9] Simkin, M.V., Roychowdhury V.P. 2005. Stochastic modeling of citation slips. Scientometrics 62(3), pp.367-384.